\documentclass{aastex61}
\usepackage{booktabs}

\received{\today}
\revised{}
\accepted{}
\submitjournal{ApJL}
\shorttitle{Fast radio bursts}
\shortauthors{Song \& Huang.}

\begin{document}

\title{The radiation mechanism of fast radio bursts}


\author{Q. W. Song}
\affil{Purple Mountain Observatory, Chinese Academy of Sciences, Nanjing, 210008, China}
\affil{Key Laboratory of Dark Matter and Space Astronomy, Nanjing 210008, China}
\author{Y. Huang}
\affil{Purple Mountain Observatory, Chinese Academy of Sciences, Nanjing, 210008, China}
\affil{Key Laboratory of Dark Matter and Space Astronomy, Nanjing 210008, China}

\author{H. Q. Feng}
\affil{Institute of space physics, Luoyang Normal University, Luoyang, 471934, China}

\author{L. Yang}
\affil{Purple Mountain Observatory, Chinese Academy of Sciences, Nanjing, 210008, China}

\author{T. H. Zhou}
\affil{Purple Mountain Observatory, Chinese Academy of Sciences, Nanjing, 210008, China}

\author{Q. Y. Luo}
\affil{Purple Mountain Observatory, Chinese Academy of Sciences, Nanjing, 210008, China}

\author{T. F. Song}
\affil{Yunnan Observatories, Chinese Academy of Sciences, Kunming 650011, China}
\affil{College of Optoelectronic Engineering, Chongqing University, Chongqing, 400044}

\author{X. F. Zhang}
\affil{Yunnan Observatories, Chinese Academy of Sciences, Kunming 650011, China}

\author{Y. Liu}
\affil{Yunnan Observatories, Chinese Academy of Sciences, Kunming 650011, China}

\author{G. L. Huang}
\affil{Purple Mountain Observatory, Chinese Academy of Sciences, Nanjing, 210008, China}

\begin{abstract}
Fast radio bursts are radio transients observed mainly around 1.5 GHz. Their peak frequency decreases at a rate of $100\sim500$MHz/s and some of them have a broader pulse with an exponentially decaying tail. Common assumptions for fast radio bursts include a dispersion effect resulting in the peak frequency drifting and a scattering effect resulting in pulse broadening. These assumptions attribute the abnormally large dispersion measure and scattering measure to the environmental medium of the host galaxy. Here we show that the radiation of fast radio bursts can be explained as an undulator radiation and the large dispersion measure can be due to a motion effect mainly from the rotation of the source which is probably variable stars. In our scenario, the pulse broadening is near-field effects and the pulse itself represents a Fresnel diffraction pattern sweeping the observer. Our work is the first analysis of properties of fast radio bursts in the context of a special mechanism of the radiation instead of a special propagation environment of the radiation.
\end{abstract}


\section{Introduction} \label{intro}
The spectrum diagram of a fast radio burst (FRB) is similar to that of a radio pulsar \citep{mas15,lor07}. Both of them show a negative frequency sweep, i.e., the higher frequency radiation arrives earlier. Observational differences between them are: (1) an FRB usually has a larger dispersion measure (DM), i.e., the peak frequency decreases slower than that of a radio pulsar and (2) for most FRBs, the frequency sweep is not repeatable while for a radio pulsar, it repeats periodically.

Interpretations of these two phenomena are different very much. While no one believes that the short pulse of a radio pulsar is a burst lasting only a few milliseconds on a neutron star, FRB is often thought of a transient burst occurring like a delta function \citep{dai16}. Taking into consideration that there are neutron stars emitting intermittently \citep{kram06} and occasionally \citep{mcl06}, and there is a repeatable FRB \citep{spi16} besides, it is not reasonable for the difference in repeatability leading to such a large discrepancy in interpretation. The difference in DM is obviously not a reason for a different timescale either. We believe that an observationally self-consistent explanation of both phenomena exists. The pulse from either of them represents a radiation cone sweeping the observer.

In this paper, we interpret the peak frequency drift of FRBs with motion effects instead of propagation effects. We achieve the goal by introducing the undulator radiation \citep{mot51} as the radiation mechanism and a model with a rotation source. We find that the mechanism is also able to explain the pulse broadening of FRBs. The main contributions of this letter are:

\begin{enumerate}
\item 	The monochromaticity variation of FRBs can be explained by an undulator radiation observed from a different angle.  We introduce this radiation mechanism in Section \ref{undul}.
\item 	To explain the frequency drift of FRBs, we introduce a model with a rotation source in section \ref{rotation}. Further analysis shows that the source of FRBs should be variable stars, which is consistent with the proposals made by \citet{mao15} and \citet{loe14}.
\item The pulse broadening  \citep{tho13} is a signature of near-field effects \citep{wal88}. We identify and discuss these effects in section \ref{pulse}.  We show that the pulse profile is a Fresnel diffraction pattern.
\end{enumerate}
Although our arguments are self-consistent, we note that other mechanism may also be responsible for the same observational phenomenon.  We propose a way to justify our explanation in Section \ref{verification}.

\section{Undulator radiation and monochromaticity} \label{undul}
The behavior of FRBs can be compared similarly to a rotating slit in optics. We begin with a qualitative discussion of two different kinds of FRBs and their association with a single slit experiment:
\begin{enumerate}
\item The intensity of FRB 150807 \citep{rav16} is the strongest among observed FRBs. Its radiation is not monochromatic. Many sparkles exist outside the burst width at each frequency that low-frequency and high-frequency emission are received simultaneously.
\item The intensity of FRB 110523 \citep{mas15} is weaker than that of FRB 150807. Its radiation is monochromatic. At any instance, the radiation is concentrated in a narrow bandwidth $\sim$ 5MHz.
\end{enumerate}

The compromise between chromaticity and intensity in the above cases has an analogy in optics. In a white light diffraction pattern from a single slit, the light is the strongest at the central maximum, but not monochrome. It is a superposition of light with different wavelengths. Near the second maximum, the light is weak, but in any position are monochrome. In radio band, a similar compromise between intensity and monochromaticity occurs when observing an undulator radiation from a different angle.

An undulator is a periodic array of dipole magnets with alternating polarity. It is commonly used to produce quasi-monochromatic synchrotron radiation with relativistic particles. The frequency of radiation in the direction of angle $\theta$ is determined by the equation \citep{jac99}:
\begin{equation}
  f\approx\frac{2{\gamma}^2}{1+{\gamma}^2\theta^2}(\frac{c}{\lambda_u})
  \label{eq1}
\end{equation}

\begin{figure}
\includegraphics[scale=.70]{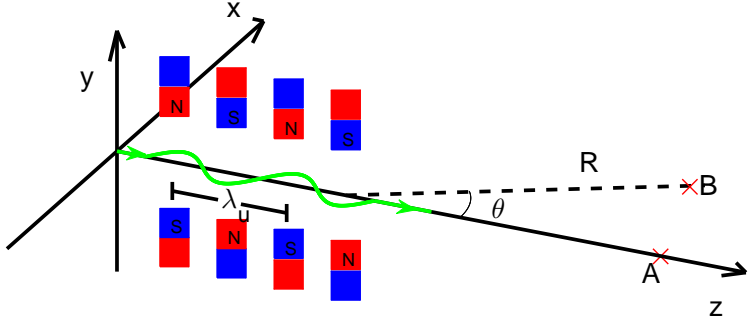}
\caption{Schematic diagram of alternating-polarity magnets (blue and red rectangles) of the period length $\lambda_u$ for an undulator. Particles are traveling along the z-axis and oscillating (green line) in a spatially periodic magnetic field. $\theta$ is the angle between the line of sight and the moving direction of particles.}
\label{fig1}
\end{figure}

Where $\gamma$ is Lorenz factor, $c$ the speed of light, $\lambda_u$ the period length of magnetic field structure. This equation is essentially describing a Doppler effect. The Doppler shift is angle dependent; the highest frequency is in the direction $\theta=0$; the frequency gets lower and lower as $\theta$ increases. The magnetic field of undulator device should be sufficiently weak so that particles moving in it are not going to be deflected away from observer's beam. The radiation received by the observer is a coherent superposition of radiation from all the periods and is monochromatic if the number of periods is large. For a typical undulator $\lambda_u \thicksim 4~cm$ and $\gamma \thicksim 10^3 $, the radiation is in X-ray wavelengths. But if we  separate those magnets as shown in Figure \ref{fig1} with thousand kilometers away from each other, the output would be in the radio band, in which FRBs are observed. The spectral width of radiation received at a given position is about $1/N$ of the observing frequency, where $N$ is the number of periods in the undulator. The dynamic spectrum width of FRBs is $\thicksim 1 MHz$; it takes $\thicksim 10^3$ magnetic periods to produce such a narrow spectrum.

Consider the electric field emitted in a weak plane undulator in Figure \ref{fig1} by a beam of electrons traveling along the z-axis. What's the difference between the emission observed by observer A and B from a different angle?

Assume the observer is at a large distance and we only consider the first harmonic:
\begin{enumerate}

\item At point A, as $\theta =0$, $f\approx\ 2\gamma^2(\frac{c}{\lambda_u})$, the emission frequency is proportional to $\gamma^2$.  $\gamma$ usually has a small spread in distribution, because particles often have an energy spread. Hence, the radiation is not monochrome, while the intensity of radiation reaches its maximum at $\theta = 0$.  This case is like FRB 150807.

\item At point B, where $\theta \gg 1/\gamma$, $f\approx\ \frac{2}{\theta^2}(\frac{c}{\lambda_u})$, the emission frequency is solely determined by $\theta$; Particles traveling at the same angle radiate at the same frequency even at different speed. This radiation is monochromatic. If it is in the near field, the peak frequency is in the off-axis direction \citep{hir84}. The intensity in large angle will be still large \citep{mos95}, but due to spectrum broadening the intensity is lower than that of A. This case is like FRB 110523.
\end{enumerate}

Not only is the chromatic property of undulator radiation similar to the diffraction of a plane wave by a narrow slit there is also an interesting analogy between the intensity distribution produced by a slit diffraction and the off-axis undulator radiation. We will discuss this in Section \ref{pulse}.

\section{Dispersion measure and  rotation of the source} \label{rotation}
Although it is not certain that the overly large DMs of FRBs partly originate from the rotation of the source itself, we wish to propose a model which assumes this. The rotation is not only important for the production of the ultrarelativistic particle beam in this model, but also for the generation of large-scale periodic magnetic structures. The rotation period predicted by this model is in the same order of magnitude of the periods of variable stars, which are found by \citet{loe14} and \citet{mao15} and proposed of the sources of FRBs.
\begin{figure}
\includegraphics[scale=.70]{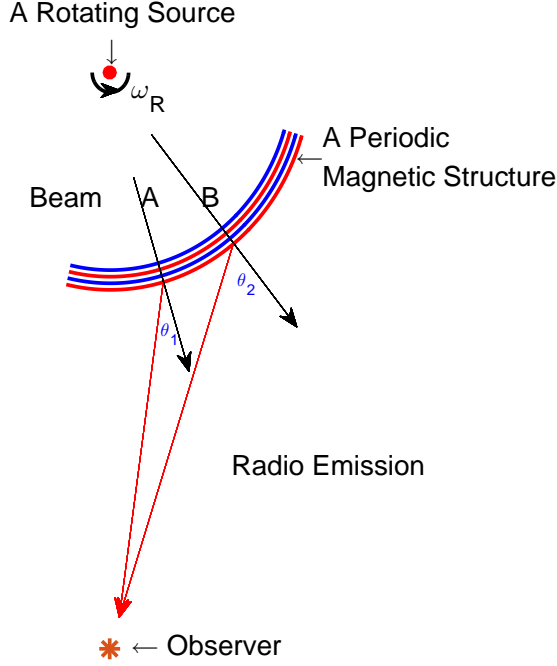}
\caption{Proposed rotation model. Beam A emits earlier than beam B. It radiates at a smaller angle and higher frequency toward the observer due to the rotation of the source.}
\label{fig2}
\end{figure}
\begin{figure}
\includegraphics[scale=.70]{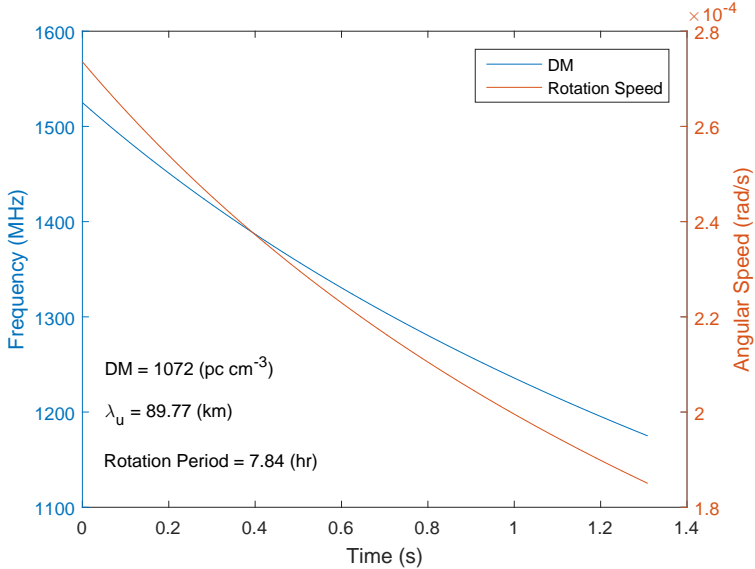}
\caption{Simulated frequency drift of FRB 110703 and corresponding angular speed, galaxy contribution already subtracted from DM. The horizontal axis refers to the time delay relative to the first pulse. Y-axis on the left is the peak frequency, on the right is the angular speed of the source.}
\label{fig3}
\end{figure}

The model is illustrated in Figure \ref{fig2}. Charged particles are accelerated to a relativistic speed in a rotating source. The outgoing direction of particles is changing at the same angular speed as that of the source. The magnetic field in acceleration region is strong. Undulator radiation can not be produced here. After traveling some distance, the particles enter a weak and axial symmetric magnetic field with many periods, as illustrated by the solid curves in Figure \ref{fig2}. The field variation of each period is not necessarily sinusoidal. Kick-like fields from magnetic discontinuities also work as long as they cause periodic deflections of particles. Undulator radiation is produced here. Generated slight earlier, beam A will produce an emission earlier than beam B. Radiating at a smaller angle ($\theta_1 < \theta_2$) to the observer, beam A will produce an emission with higher frequency than that of beam B. This is how the time delay from the high-frequency radiation to the low-frequency is produced. Note this is an ultra-relativistic case; we must distinguish the emitter time from observer time. Because particles are chasing the wave front, a pulse lasting one second is produced by the source in $\sim 2\gamma^2$ seconds. However, a simple frequency decreasing due to the uniform rotation is not enough to produce a dispersive like time delay. Our simulation of FRB 110703 (shown in Figure \ref{fig3}) shows that an angular acceleration speed should exist. This type of none-uniform angular speed usually exists in orbital motion.

With the approximation $f\approx\ \frac{2}{\theta^2}(\frac{c}{\lambda_u})$, an estimator of the rotation period of the source can be given by:

  \begin{equation}
 \hat{P}=\frac{2\pi}{\bar{\omega}}=4148.808s\times\frac{DM}{(pc~cm^{-3})}\times\pi\sqrt\frac{2\lambda_u}{c}(\sqrt\frac{1}{\nu_l}+\sqrt\frac{1}{\nu_u})(\frac{1}{\nu_l}+\frac{1}{\nu_u})
 \label{eq2}
\end{equation}

Where   ${\nu_l}$ denotes the lower limit of the observation frequency in the unit of MHz,  ${\nu_u}$ the upper limit, $\lambda_u$ the length of a magnetic period.    If DM is 1103 $pc~cm^{-3}$(as of FRB 110703) and $\lambda_u$ is between $10~km$ and $10000~km$, then the rotation period of the source is between 2 hours and 3.4 days. This range is coincident with the period range of the variable stars observed by \citet{loe14} and \citet{mao15}. They found three variable stars near the location of FRB 110703, FRB 110627 and FRB 010621, two of which are low mass contact binaries, the other one is a slowly pulsating B-star \citep{de2007}. The periods of these stars are from 7.8 hours to 2.5 days. They proposed that these variable stars are the sources of the corresponding FRBs. Without a priori knowledge of $\lambda_u$, we are not able to verify their proposal by estimating period from $\lambda_u$, but with a known period of the star, we can estimate $\lambda_u$  from Equation \ref{eq2}. The $\lambda_u$ of FRB 110703 is about $90~km$. The observation angle $\theta$ increases from ~7.22 arcmin to 7.56 arcmin.

This model can explain why the DMs of FRBs are so large. The reason is that the rotation also contributes to the time delay between the high-frequency and low-frequency radiation. However, it can not explain why the time delay is dispersive like. A selection effect seems to exist that only when the orbital motion produces a time delay $\sim\nu^{-2}$, the event is identified as an FRB.

\section{Pulse broadening and near field effects} \label{pulse}
The pulse shape of FRBs carries the most important clues to the radiation process. Several bursts of the repeating fast radio burst FRB 121102 \citep{cha17,spi16} have multi peaks. FRB 130729 \citep{cha16} has a right peak lower than the left one. And there is a variety of pulses from different FRBs with different asymmetries in shapes. From the spectrum point of view, the dynamic spectra of these FRBs are variable.  We will never be able to give an explanation to them unless we find a mechanism capable of producing a diversity of spectra in this radio band.

\begin{figure}
\includegraphics[scale=.70]{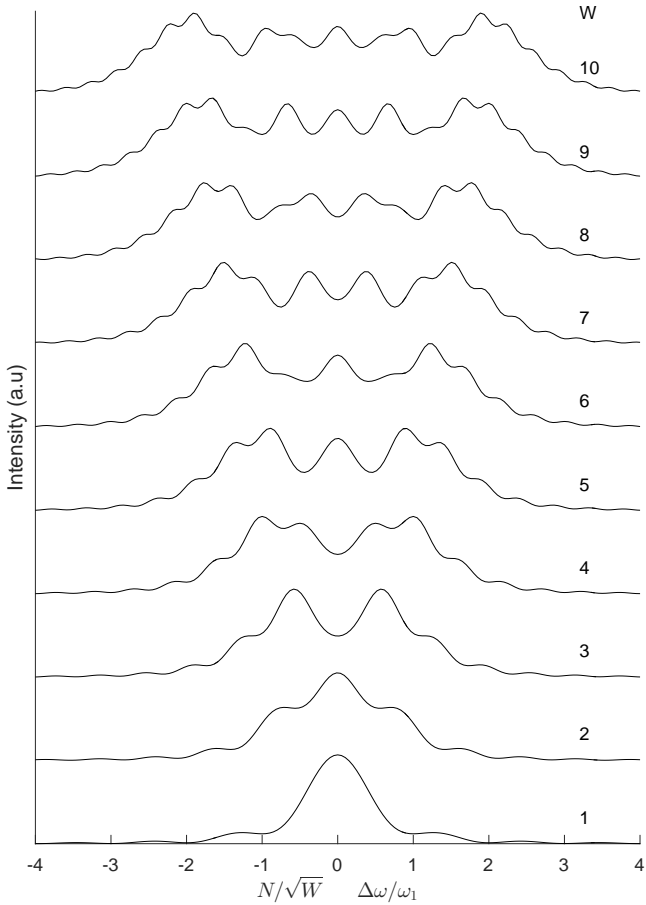}
\caption{Spectral distributions around the central frequency $\omega_1$ in the near field case reduced to an intensity distribution of a Fresnel diffraction pattern from a single slit. The width of the slit is $2\sqrt{W}$. The horizontal axis denotes the reduced frequency deviation from the central frequency $\omega_1$. The  vertical axis is the intensity in arbitrary unit.}
\label{fig4}
\end{figure}

The suggestive clue regarding the radiation mechanism comes from the double-peaked pulse of FRB 121002 \citep{cha16}. Its pulse contains two peaks. The dynamic spectrum also consists of two peaks separated each other from the central frequency. For undulator radiation, such a splitting of spectrum suggests near-field effects \citep{wal88}. Walker modeled the effects analytically and numerically. The double-peaked spectrum appears when the light from different parts of the source develop a path length difference of $\sim$ 3/4 wavelength (W=3 in Figure \ref{fig4}) at the observer position.

Near-field effects will broaden the spectrum, split the spectrum into several peaks. It will affect both the spectral width and the angular spread of spectrum. Our explanation is the first work applies it to explain an astrophysical phenomenon. We prefer to give a brief explanation to this effect in the context of the radio radiation produced in stellar magnetic structures. Near field effects are described by the parameter:
\begin{equation}
  W=\frac{L^2\theta^2}{2\lambda D}
\end{equation}
in which $L$ is the total length of the periodic magnetic structure. $\lambda$ is the wavelength of radiation. $D$ is the distance from the source to observer and $\theta$ is the observation angle. $L\theta$ is the apparent size of the magnetic structure. It is equivalent to the width of a single slit. When the width ($L\theta$) is larger than the geometric average of wavelength and distance, near-field effects will occur. The observer could not be thought of at infinity and the electromagnetic wave should be treated as a spherical wave.  If $W=1$, a $\pi/2$  phase difference is among the output radiation.

Walker's analysis in the phase space shows that both spectrum and angular distribution in a near field case can be reduced to a dimensionless intensity distribution of Fresnel diffraction. We replot his result in Figure \ref{fig4}, where $\omega_1$ is the central frequency given by Equation \ref{eq1} for a given observation angle $\theta$, $\triangle\omega$ is the small deviation from $\omega_1$ . For the spectrum distribution, the dimensionless $X$ coordinate is $\frac{N}{\sqrt{W}}\frac{\triangle\omega}{\omega_1}$, it is equal to the distance from a screen point to the center of a single slit diffraction pattern. The reduced width of a slit is $2\sqrt{W}$. Different $W$ means different slit width. The double-peaked spectrum of FRB 121002 is equivalent to an interference fringe we see in a slit diffraction pattern when $W=3$.

All curves in Figure \ref{fig4} are symmetric because Walker's analysis in phase space didn't consider the angular variation of intensity. With this simplification, the angular distribution and the spectrum distribution are identical, i.e., the spectral shape in Figure \ref{fig4} is also the shape of the corresponding pulse, which is true for FRB 121002. His numeric simulation taking the variation of intensity into account shows that the spectrum is generally asymmetric. It depends on the observation angle and harmonics. The pulse shapes of  the No. 5, 7 and 10 burst of FRB 121102 \citep{spi16} are comparable to some angular distributions given by the numeric simulation for $W>4$.

With large angle approximation, $W=\frac{N^2\lambda_u}{D}$. where $N$ is the number of magnetic periods. $W$ is mainly determined by $N$, because it is proportional to the square of $N$. If $N$ is sufficiently large, we will observe near-field effects even from the source at a cosmological distance. If the repeating FRB 121102 is really at a distance of $972~MPc$ \citep{ten17}. For $W=4$ and $\lambda_u=90~km$, $N$ is $\sim 3.65\times10^{10}$. The size of the source is $\sim 0.35$ light year.

Near field effects provide a competing interpretation to the pulse profile of FRBs, other than the scattering broadening. No matter which process is really behind FRB 121102, the mathematical form of the process likely includes a Fresnel integral. These effects also add a new mechanism to explaining the spectral structure in  the GHz and MHz band astronomical observation.

\section{Discussion} \label{discussion}
\subsection{verification of the mechanism} \label{verification}

A simple way to verify the mechanism proposed by us is to check whether there is a positive frequency sweep ahead of the negative frequency sweep. The intensity distribution of undulator radiation is symmetric to the beam \citep{mos95}. We will be swept firstly by a cone whose low-frequency radiation reaches us first.

\subsection{density modulation of the beam}
Not only a periodic magnetic field distribution  can increase the spectral flux, but a periodic density distribution of particles can do. The latter one is called a density modulation. We'll use the sparkle structure observed in the strongest FRB 150807 \citep{rav16} as an example to explain the idea.

 The observation of  FRB 150807 shows many 100 kHz sparkles. For a GHz radiation, 100 kHz bandwidth means the coherence length is about $\sim 10^{4}$ $\lambda$. Two processes will produce the sparkles:
 \begin{enumerate}
 \item A beam of particles passes through  a magnetic structure with $\sim 10^{4}$ periods. This is a process in a usual undulator device, which we've already discussed.
 \item A beam of particles, divided into $10^4$ bunches and separated each other by $\lambda$, passes through a magnetic structure with only a few periods. This is a process with a density modulation involved, which hasn't been discussed.
 \end{enumerate}
 The bandwidths of the radiation produced by  both processes above are equal to each other. The output energy of the latter one will greatly increase by the order of particle number. The latter scheme is used in a free-electron laser (FEL) experiment to increase pulse energy by $10^{10}$ times \citep{Gel10}. The density modulation produced by a natural process may be much less perfect than the modulation achieved by an FEL experiment. But as long as the modulation exist, it will increase the spectral flux. Some giant pulses from the Crab pulsar can exceed 2MJy \citep{Han07}. The intensity of the FRB 150807 sparkles also exceeds 1 KJy. Such a variation of intensity by orders implies a density modulation exist. The magnitudes of events are decided by how much particles are modulated and involved in the coherent process.

 The low occurrence rate of FRBs is similar to the giant pulses from young pulsars. They are conjectured to be the same things \citep{kea12,kat16}. From the density modulation point of view, the low occurrence rate reflects the difficulty for a natural process in binary stars or a neutron star to produce a perfectly modulated beam like the one produced in  a man-made FEL. These two events are not necessarily the same things. Their requirements for a perfectly density modulated beam are same to each other. So both the low occurrence rate and the uncommonly high intensity of radiation support the existence of a density modulation of the beam.

\subsection{periodic magnetic structure}
The large-scale periodic magnetic structure will be easily destroyed by turbulence.  Then what is the periodic magnetic structure of our model in practice? Could it be generated in stellar space?    In our opinion,  for a magnetic structure with $\lambda_u \sim 90~km$,  even if it is  around the Sun, the closest star to us, we are not able to observe it with current technology and can't prove the existence of the structure by observation. On the contrary, the mechanism proposed in this letter provides a method to infer it. The information we can tell from the polarization of radiation is:
\begin{enumerate}
\item Some FRBs are produced in  planar magnetic structures so that their polarizations are linear. In interplanetary space, several sinusoidal magnetic periods are  usually ahead of a shock \citep{Feng08} driven by a magnetic cloud. The magnetic field there is planar. Combining with a density modulated beam, a monochromatic and linear polarized radiation can be produced.
\item Some periodic structures are not planar. So the radiation produced in them are not polarized. They may be periodically distributed discontinuities or plasma oscillations frequently generated everywhere.
\end{enumerate}
The magnetic structure is not going to be too long if only a density modulation exists.

%

\acknowledgements
This research is supported by the Opening Project of Key Laboratory of Astronomical Optics \& Technology, Nanjing Institute of Astronomical Optics \& Technology, Chinese Academy of Sciences. It is also supported by NSFC 11533009, U1631135 and 11203083.



\end{document}